\documentclass{article}
\usepackage{spconf,amsmath,amssymb,graphicx,booktabs}
\usepackage{enumitem}
\usepackage{balance}
\usepackage[hypertexnames=false]{hyperref}
\hypersetup{hidelinks}
\newcommand{\method}{TRACE}
\newcommand{\state}[1]{\emph{#1}}
\newcommand{\AttrModelCount}{8}
\newcommand{\AttrConditionCount}{27,648}
\newcommand{\AttrCorrectCount}{18,911}

\newcommand{\Confused}{2.7\%}

\newcommand{\Ungrounded}{3.6\%}

\newcommand{\RankFail}{6.3\%}
\newcommand{\RankFailCI}{4.9--7.8\%}
\newcommand{\RankFailCount}{1,184}

\newcommand{\TargetOnlyMissRank}{42.5\%}

\newcommand{\FlipFlagAll}{30.8\%}

\newcommand{\CICoverRank}{89.1\%}

\newcommand{\FlipFlagOrderedAll}{24.5\%}

\newcommand{\AttrMixtureCount}{3,456}

\newcommand{\ThreshShiftZeroShare}{6.3\%}
\newcommand{\ThreshShiftZeroN}{18,911}

\newcommand{\ThreshShiftMinShare}{4.6\%}
\newcommand{\ThreshShiftMaxShare}{8.5\%}

\newcommand{\SnrRankFailHigh}{4.9\%}

\newcommand{\SnrRankFailMid}{6.3\%}

\newcommand{\SnrRankFailLow}{8.4\%}

\newcommand{\SnrRankFailDelta}{3.5}
\newcommand{\SnrRankFailDeltaCI}{2.7,\,4.3}
\newcommand{\SnrRankFailIncreasingModels}{8}
\newcommand{\SnrRankFailMinModelLift}{1.9}

\newcommand{\SnrAccHigh}{75.7\%}

\newcommand{\SnrAccLow}{60.8\%}

\newcommand{\CatCount}{49}

\newcommand{\ExModel}{Qwen3-Omni-30B}

\newcommand{\SubstUnits}{23,040}
\newcommand{\SubstConditions}{8,064}
\newcommand{\SubstClips}{160}
\newcommand{\SubstRoles}{56}
\newcommand{\SubstModels}{8}
\newcommand{\SubstModelsOrdered}{8}

\newcommand{\SubstGapFrozen}{+20.9}

\newcommand{\SubstGapSubst}{+15.7}
\newcommand{\SubstGapSubstCI}{[+10.1,\,+21.8]}
\newcommand{\SubstDiD}{-5.3}
\newcommand{\SubstDiDCI}{[-9.3,\,-1.9]}

\newcommand{\SubstGapFrozenBoth}{+22.6}
\newcommand{\SubstGapSubstBoth}{+20.0}
\newcommand{\SubstGapSubstBothCI}{[+13.3,\,+27.4]}
\newcommand{\SubstRoleRho}{+0.69}
\newcommand{\SubstRoleN}{56}

\newcommand{\AHACommonCorrectCount}{2,466}

\newcommand{\ReplRankFail}{13.8\%}

\newcommand{\ReplRoleCount}{192}

\newcommand{\ReplFrozenMatched}{10.8\%}

\newcommand{\ReplFrozenMatchedDenom}{5,220}
\newcommand{\ReplRepairMatched}{10.6\%}

\newcommand{\ReplMatchedDelta}{-0.3}
\newcommand{\ReplMatchedDeltaCI}{[-3.0,\, +2.4]}

\newcommand{\ProbeLoudnessRankFail}{45.0\%}

\newcommand{\MatchSharedAcoustics}{64}

\newcommand{\MatchPerModelAbsent}{128}
\newcommand{\MatchAbsentPooled}{512}
\newcommand{\MatchPaperTPR}{55.2\%}
\newcommand{\MatchPaperFPR}{23.6\%}

\newcommand{\HDesignExactZero}{86}

\newcommand{\HDesignBandFiveHundredths}{5.5\%}

\newcommand{\CatGap}{+21.5}
\newcommand{\CatGapCI}{[15.7,\,27.8]}
\newcommand{\CatRepGap}{+25.9}
\newcommand{\CatRepGapCI}{[17.3,\,36.1]}
\newcommand{\CatRepModels}{3}

\newcommand{\CatRho}{+0.79}
\newcommand{\CatShared}{49}
\newcommand{\CatMarginLo}{+17.1}
\newcommand{\CatMarginHi}{+26.6}

\newcommand{\CatQWOne}{22.3\%}

\newcommand{\CatQWFive}{0.8\%}

\newcommand{\UtilConflictN}{173}
\newcommand{\UtilRepairableN}{420}
\newcommand{\UtilTotalN}{593}

\newcommand{\UtilConflictRedrawN}{566}
\newcommand{\UtilRepairableRedrawN}{1,724}

\newcommand{\UtilConflictFail}{35.5\%}
\newcommand{\UtilRepairableFail}{11.7\%}
\newcommand{\UtilRawDelta}{+23.9}

\newcommand{\UtilMatchedDelta}{+13.5}
\newcommand{\UtilMatchedDeltaCI}{+4.1,\,+22.7}
\newcommand{\UtilModelsPositive}{3}
\newcommand{\UtilMatchedUsed}{434}
\newcommand{\UtilCellCount}{33}

\newcommand{\UtilSameMapDelta}{+13.7}
\newcommand{\UtilSameMapDeltaCI}{+2.7,\,+25.2}
\newcommand{\UtilSameMapUsed}{338}

\newcommand{\MapAFail}{4.6\%}
\newcommand{\MapACommonN}{2,355}

\newcommand{\MapAGradient}{+2.0}

\newcommand{\MapBFail}{7.6\%}
\newcommand{\MapBCommonN}{3,111}

\newcommand{\MapBGradient}{+4.6}
\newcommand{\MapBalanceA}{50.0\%}

\newcommand{\InterfacePairN}{6,144}
\newcommand{\InterfaceBothN}{2,670}

\newcommand{\InterfaceVerdictAgreeBoth}{92.4\%}

\newcommand{\InterfaceKappaRfBoth}{0.71}
\newcommand{\InterfaceEnrich}{7.3$\times$}

\newcommand{\InterfacePearson}{+0.91}

\title{Right Answers, Inconsistent Evidence:\\
Source-Level Counterfactual Auditing\\
of Large Audio-Language Models}
\name{Yi Zhang$^{1}$, Yi Li$^{2}$, Hongwei Du$^{1}$\sthanks{Corresponding
author: hongwei.du@ieee.org.}}
\address{$^{1}$School of Computer Science and Technology,
Harbin Institute of Technology, Shenzhen, China\\
$^{2}$University of Windsor, Windsor, Canada}

\begin{document}
\ninept
\maketitle

\begin{abstract}
A large audio-language model (LALM) can answer ``is a cat present?'' about a
mixture correctly, and its confidence can drop when the cat is removed---yet
this does not show that it can tell the cat's presence from its absence. We
introduce \method{}. It deletes the queried source and each rival in turn and
asks whether every view still containing that sound outscores the view with it
removed. Across \AttrModelCount{} LALMs and \AttrMixtureCount{} controlled
mixtures, \RankFail{} of correct answers fail this ordering: the target-absent
view scores at or above some target-present view, a conflict no shared
threshold repairs. A target-only score check misses \TargetOnlyMissRank{} of
them. The share rises from \SnrRankFailHigh{} to \SnrRankFailLow{} as target
SNR falls from $+5$ to $-5$~dB, even on units correct at every level, and
survives loudness and re-pairing controls. Accuracy certifies the answer;
\method{} audits whether it orders presence above absence. Code and data:
\url{https://github.com/yikise/TRACE}.
\end{abstract}


\section{Introduction}
\label{sec:intro}

Large audio-language models (LALMs) \cite{chu2023qwen,tang2023salmonn,
gong2023listen,ghosh2024gama,chu2024qwen2audio,xu2025qwen25omni,
xu2025qwen3omni,goel2025audioflamingo3} answer natural-language questions
about arbitrary recordings. Real recordings are rarely a single sound: a cat,
a car horn, and laughter overlap in one clip, and the model must report on one
of them. A correct ``yes'' may therefore rest on the cat itself, or on a
scene that merely contains one, and accuracy cannot tell the two apart. That
distinction matters wherever a presence judgment drives action: a model
reporting the right event for the wrong reason will not fail gracefully when
the scene shifts.

The checks that guard against this failure share a blind spot. Hallucination
benchmarks ask whether absent events are reported
\cite{li2023pope,kuan2024understanding,kuan2025trulyhear,cheng2025ahabench,zhao2026halluaudio}.
Target-removal tests ask whether the answer about a deleted event flips, or
whether answers about other events stay stable under removal
\cite{kuan2025trulyhear,moon2024beaf}. Each reads the counterfactual
\emph{answer}. None compares how the model \emph{scores} the queried sound
against the views that still contain it. A target-only score check misses the
cases where a rival's removal lowers the score further. Grading the answer on
\emph{every} deletion view does catch those samples as counterfactually wrong,
but reports one undifferentiated error: it cannot say whether the ordering of
evidence broke, or the decision threshold merely fell inside a consistent
ordering. That decomposition is what is missing.

\begin{figure*}[t]
  \centering
  \includegraphics[width=0.90\textwidth]{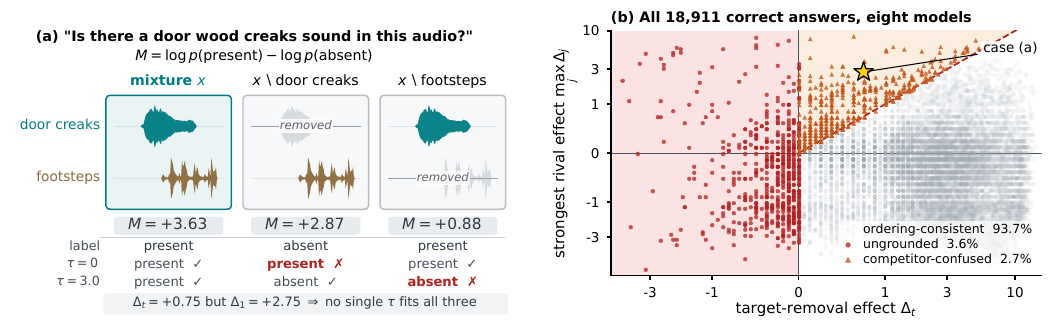}
  \caption{Overview of \method{}. (a)~A mixture (door wood
  creaks + footsteps; \ExModel{}, $0$~dB, partial overlap). The target-only
  check passes, yet removing the rival lowers the score further, so no shared
  threshold fits all views. (b)~All \AttrCorrectCount{} correct answers by
  $\Delta_t$ and $\max_j\Delta_j$ (symlog): $\Delta_t\le 0$ ungrounded, on or
  above the dashed line competitor-confused, the rest ordering-consistent.
  Star: case in (a).}
  \label{fig:overview}
\end{figure*}

The criterion we adopt is ordinal. A model that has registered the queried
sound should score every view that still contains it above the view with that
sound removed. A single decision threshold can repair a misread answer, but no
threshold shared across the views can repair an ordering conflict, so the two
call for different remedies, and $H\le0$ names the latter. We introduce
\method{} (Target-and-Rival Ablation for Counterfactual Evidence) to find such
cases at scale: it deletes the queried source and every rival in turn, re-scores
the same question on each view, and composes the removal effects into the
ranking statistic $H$, positive only when every target-present view outscores
the target-absent view. Deleting the target turns its label to absent while
deleting a rival leaves it intact, so \method{} needs only the answer logits a
frozen model already produces over a known, deletable source set---no generated
text, no retraining, and no new recordings.
Fig.~\ref{fig:overview}a shows the failure on a measured mixture: \ExModel{},
the panel's most accurate model, answers a door-creak question correctly and
passes the target-only check, yet deleting the rival footsteps lowers the margin
further than deleting the door creak itself.

We evaluate \method{} on \AttrModelCount{} open LALMs over
\AttrMixtureCount{} controlled ESC-50 mixtures crossing target SNR,
distractor count, and temporal overlap, with every view's presence label fixed
by construction. Among \AttrCorrectCount{} correct answers, \RankFail{} are
ranking failures. Our contributions are summarized as follows:
\begin{itemize}[nosep,leftmargin=11pt]
  \item We introduce \method{}, a source-level counterfactual test, and its
  ranking statistic $H$, which isolates the answers no shared threshold orders
  correctly.
  \item We show that these failures hide inside correct answers: deleting the
  target still lowers the score, only less than deleting a rival does, so a
  target-only check misses \TargetOnlyMissRank{} of them and the full answer
  check cannot separate them from threshold-repairable errors.
  \item We establish that the conflicts are competition-driven and
  reproducible: they grow with acoustic competition, survive loudness and
  re-pairing controls, concentrate on specific source events, and hold a
  positive high--low risk gap under within-category replacement.
\end{itemize}

\section{Related Work}
\label{sec:related}

\textbf{Evaluating audio-language models.}
AIR-Bench, AudioBench, and MMAU assess broad comprehension
\cite{yang2024air,wang2025audiobench,sakshi2025mmau}, temporal benchmarks add
reasoning across clips and time \cite{chen2024beyondsingle,xie2025audiotime},
and audio-text models match on nouns and verbs rather than the full sentence
\cite{wu2023naturallanguage}.

\textbf{Hallucination and counterfactual tests.}
Whether a correct prediction rests on the right evidence predates the current
models \cite{ross2017right}. Discriminative yes/no probes
were introduced for object hallucination in
vision \cite{li2023pope} and adapted to audio
\cite{kuan2024understanding,kuan2025trulyhear,cheng2025ahabench,zhao2026halluaudio}; AHA aligns a
model with counterfactual hard negatives \cite{chen2025aha}. Mousavi et al.\
intervene on both the audio and the generated chain of thought to test whether
a reasoner's explanation is grounded in what it heard
\cite{mousavi2026faithfulness}. BEAF scores
before--after answer changes under object removal and tracks whether answers
about objects never removed stay stable \cite{moon2024beaf}, a design also
used for sound events \cite{kuan2025trulyhear}.

\textbf{Perturbation-based attribution.}
LIME fits local surrogates to perturbed inputs \cite{ribeiro2016lime} and
audioLIME toggles source-separated components \cite{haunschmid2020audiolime}.
\method{} instead uses ground-truth stems, so each view carries an exact
presence label.

\section{Method}
\label{sec:method}

\subsection{Source-matched views}
Let $x$ be a mixture of known stems, $t$ the queried source, and $D$ the
set of rival sources. For one fixed presence question $q$ with permitted
answers $y_p$ (present) and $y_a$ (absent), we define the margin
\begin{equation}
M(x,q)=\log p(y_p\mid x,q)-\log p(y_a\mid x,q),
\label{eq:margin}
\end{equation}
and read the answer as present when $M>0$. We score three kinds of views
with the same $q$: the mixture gives $m_f=M(x,q)$, the target-removed view
gives $m_t=M(x\setminus t,q)$ with label absent, and each rival-removed view
gives $m_j=M(x\setminus j,q)$ for $j\in D$ with label present. Each deletion
drops one stem and leaves the gains and offsets of the others unchanged, so
the removal effects are
\begin{equation}
\Delta_t=m_f-m_t,\qquad \Delta_j=m_f-m_j .
\label{eq:effects}
\end{equation}
A positive $\Delta_t$ means the target supports the answer, a positive
$\Delta_j$ that rival $j$ does, and a negative $\Delta_j$ that removing $j$
raises the target's margin.

\subsection{Ranking failures}
For a correct mixture answer ($m_f>0$), we call the answer a \emph{ranking
failure} when
\begin{equation}
H=\Delta_t-\max(0,\max_j\Delta_j)
=\min\!\left(m_f,\min_jm_j\right)-m_t\;\le\;0,
\label{eq:states}
\end{equation}
that is, when the target-removal effect is non-positive, or a rival's
removal costs at least as much support as removing the target. Two
sub-cases make the failure interpretable. If
$\Delta_t\le0$ (\state{ungrounded}), the model reports the target with
equal or greater confidence after the target is deleted, so even the
target-only check fails; if $\Delta_t>0$ but $\max_j\Delta_j\ge\Delta_t$
(\state{competitor-confused}), the target-only check passes while the
target-absent view outscores a target-present view. These labels describe score
behavior, not identified internal mechanisms. Answers with
$H>0$ are \emph{ordering-consistent}: a rival's removal may still move
the score more than the target's without breaking the ordering.

The reading is ordinal and stable rather than knife-edged. $H>0$ holds iff the
target-absent view scores below every target-present view---equivalently, iff a
shared local threshold $m_t\le\tau<\min(m_f,\min_jm_j)$ classifies every view
correctly---while $H\le0$ certifies an ordering conflict no shared threshold
repairs, and its sign is invariant to any strictly increasing rescaling of $M$
shared across views. Only \HDesignExactZero{} of the \AttrCorrectCount{}
correct answers sit at an exact tie, and counting every answer with
$|H|\le0.05$ as ordered lowers the share only to \HDesignBandFiveHundredths{}.
$H>0$ certifies local separability by one threshold only, not that the model
identifies the queried source as a listener would.

Shares are conditional on a correct mixture answer, the population in which a
correct answer can still hide a broken ordering. That condition is a set
definition, not the source of the effect: moving the cut $\tau$ in $m_f>\tau$
over $\tau\in[-1,2]$ only changes which answers enter, with the share between
\ThreshShiftMinShare{} and \ThreshShiftMaxShare{} (at $\tau=0$,
\ThreshShiftZeroShare{} of \ThreshShiftZeroN{}). Sec.~\ref{sec:states}
reports the matched present- and absent-question rates.

\subsection{Relation to existing checks}
Prior removal tests grade the counterfactual answer rather than the scores
behind it, so we compare them against the ordering test on the same correct
answers.
\begin{table}[t]
\centering
\caption{Decomposition of counterfactual checks on the same
\AttrCorrectCount{} correct mixture answers (\%). All columns share that
denominator and Flagged $=$ Conflict $+$ Separable throughout. The target-only
and ordering rows have Conflict $=$
Flagged by construction: set identities, not measured detection rates.}
\label{tab:checks}
\small
\setlength{\tabcolsep}{4.2pt}
\renewcommand{\arraystretch}{0.97}
\begin{tabular}{lrrr}
\toprule
Check & Flagged & Conflict & Separable \\
\midrule
Target-only score & 3.6 & 3.6 & 0.0 \\
Target-removed answer & 27.3 & 5.6 & 21.7 \\
Answer flip & 30.8 & 6.3 & 24.5 \\
\method{} (ordering) & 6.3 & 6.3 & 0.0 \\
\bottomrule
\end{tabular}

\end{table}

Table~\ref{tab:checks} decomposes the same \AttrCorrectCount{} correct
answers into overlapping alarm sets: the target-only score check alarms when
$\Delta_t\le0$, the target-removed answer check when $m_t>0$, and the
answer-flip rule when $m_t>0$ or $\min_jm_j\le0$. The flip rule alarms on
\FlipFlagAll{}, of which \RankFail{} carry an ordering conflict and
\FlipFlagOrderedAll{} are locally threshold-separable; \method{} isolates the
ordering conflicts. The two
target-removal families miss different amounts: the target-only \emph{score}
check never queries the rival axis and misses \TargetOnlyMissRank{}, while
the target-removed \emph{answer} check covers \CICoverRank{} and still misses
the remainder. Separability is local to each mixture's own view set, not one
global threshold.

\section{Experimental Setup}
\label{sec:setup}

\subsection{Controlled mixtures}
We source events from ESC-50 \cite{piczak2015esc50} (5-second clips, 50
classes), drawing 192 clips spanning 49 classes and arranging them into 64
three-source tuples with distinct
labels. Rotating the queried source through each tuple gives 192 target
roles. Each role is crossed with target SNR ($-5$, $0$, $+5$~dB),
temporal overlap (full or partial), and distractor count (one or two);
with one distractor, each rival is used in turn. This yields 3,456
mixture conditions per model, \AttrConditionCount{} model--mixture
evaluations across the panel. The target gain is fixed and the included
distractors are jointly scaled to set the target-to-distractor SNR, computed
from whole-clip RMS over each rendered 4-second crop. Full-overlap
sources share an onset at $1$~s; partial-overlap distractors lead or lag the
target by $0.75$~s; mixtures are $4$~s long.
Labels follow the ESC-50 category annotations. Two listeners, blind to source
role and model outcome, reviewed all 192 panel
crops against the exact audio we render and confirmed the declared event with
no other nameable event audible. Every tuple uses three distinct categories,
so each deletion removes exactly one label and every removal view's presence
label is fixed by construction. Re-paired backgrounds draw both distractors
from the panel's 49-category source list, never repeating the queried
category. All 11,520 views reuse the same 192 hash-verified clips.

\subsection{Models and scoring}
We evaluate Qwen2-Audio-7B \cite{chu2024qwen2audio}, Qwen2.5-Omni-3B and
7B \cite{xu2025qwen25omni}, Qwen3-Omni-30B \cite{xu2025qwen3omni},
Step-Audio 2 Mini \cite{wu2025stepaudio2}, MiMo-Audio-7B
\cite{xiaomi2025mimoaudio}, Audio Flamingo 3 \cite{goel2025audioflamingo3},
and the AHA LoRA adapter on Qwen2.5-Omni-7B \cite{chen2025aha}, which
provides a same-base training contrast. Questions are discriminative with
lettered answers (A/B): ``Is there a \emph{category} sound in this audio?
Answer with A for yes or B for no.'', with the mapping balanced
\MapBalanceA{} across the panel and fixed within each role across deletion
views, SNR levels, and models. We read $M$ from
the two answer logits, so every answer is parsed. The ranking-failure share is
\MapAFail{} under one mapping and \MapBFail{} under the other, and the
competition gradient is positive under both (\MapAGradient{} on
\MapACommonN{} units, \MapBGradient{} on \MapBCommonN{}); these rates use
different roles. Interface robustness is measured on a separate query-swap
panel re-rendering every role from identical audio under four wordings and
both mappings, so only the interface varies: paired verdict agreement is
\InterfaceVerdictAgreeBoth{} on the \InterfaceBothN{} jointly correct pairs,
with $H$ correlated \InterfacePearson{} over all \InterfacePairN{}. On those
same pairs failure verdicts agree beyond what each mapping's own rate predicts
($\kappa=\InterfaceKappaRfBoth{}$, \InterfaceEnrich{} the overlap of
independent verdicts). A separate absent-event panel poses
\MatchAbsentPooled{}
conditions about events absent from every stem, built from
\MatchSharedAcoustics{} full-overlap two-distractor mixtures crossed with
four wordings and both mappings; the matched comparison in
Sec.~\ref{sec:states} uses the canonical wording, giving
\MatchPerModelAbsent{} absent questions per model.

\subsection{Controls and statistics}
Deleting a stem lowers total energy, so we recompute all effects on an
RMS-matched control that rescales every removal view to its mixture's RMS,
with a shared peak-limiting gain when needed. Confidence intervals
are 95\% percentile intervals from 10,000 bootstrap draws over the 64 source
tuples (42 in the substitution panel), or the 192 target roles where an
analysis re-pairs backgrounds, keeping all models and conditions within a
sampled cluster, and rates
are ratios of pooled counts recomputed in every draw. Stress-axis
comparisons are paired: only evaluation units (model, recording, target role)
correct at every level of the varied factor enter, so the evaluated answers
are identical across levels.

Every view is scored from the next-token log-probability of its two lettered
answers under each model's own prompt format, without generating text, so $M$
in Eq.~\ref{eq:margin} is deterministic given the checkpoint. The main panel
poses every question under the canonical wording with one mapping per role; the
absent-event panel and the paired audit cross the wordings and mappings. Every
prompt closes with an answer key that allows A or B and the model's own
``Answer:'' prefix. Under
every checkpoint each letter is a single token appended to the prompt, so its
score is read at the position that predicts the appended token rather than as a
continuation of preceding text. Audio is resampled to 16~kHz (24~kHz for
MiMo-Audio), models run in bfloat16, the bootstrap uses a fixed seed, and every
reported difference comes from unrounded shares.
\section{Results}
\label{sec:results}

\subsection{Ranking failures hide inside correct answers}
\label{sec:states}

\begin{table}[t]
\centering
\caption{Attribution of correct mixture answers (\%). Acc.\ is mixture accuracy
on present-event questions; Conf.\ and Ungr.\ split the ranking failures into
the two sub-cases, and RFail ($H\le0$) comes from the same counts, so it can
differ from their rounded sum. Correct-answer subsets differ across models, so
these rates are unmatched.}
\label{tab:main}
\small
\setlength{\tabcolsep}{2.6pt}
\renewcommand{\arraystretch}{0.97}
\begin{tabular}{lrrrr}
\toprule
Model & Acc. & Conf. & Ungr. & RFail \\
\midrule
Audio Flamingo 3 & 60.4 & 1.3 & 2.2 & 3.5 \\
MiMo-Audio-7B & 46.3 & 4.5 & 1.4 & 5.9 \\
Qwen2-Audio-7B & 74.4 & 3.5 & 5.6 & 9.1 \\
Qwen2.5-Omni-3B & 62.9 & 1.9 & 2.3 & 4.2 \\
Qwen2.5-Omni-7B & 78.6 & 2.1 & 1.4 & 3.6 \\
\quad + AHA (LoRA) & 72.0 & 1.9 & 1.3 & 3.2 \\
Qwen3-Omni-30B & 84.3 & 2.2 & 2.7 & 4.9 \\
Step-Audio 2 Mini & 68.4 & 4.2 & 11.3 & 15.6 \\
\midrule
Pooled & 68.4 & 2.7 & 3.6 & 6.3 \\
\bottomrule
\end{tabular}

\end{table}

Table~\ref{tab:main} attributes all
\AttrCorrectCount{} correct mixture answers. Pooled over models, \RankFail{}
are ranking failures (\RankFailCount{}, 95\% CI \RankFailCI{}), split
into \Confused{} competitor-confused cases, where the target-only check passes
but a rival's removal costs more, and \Ungrounded{} ungrounded cases, where it
fails outright. Every model contributes failures, and their number does not
follow accuracy: Qwen2-Audio-7B is third in accuracy yet second in failures.
The target-only score check misses \TargetOnlyMissRank{} of these conflicts
(Table~\ref{tab:checks}).

A separate absent-event panel asks the same
mixtures about categories no stem contains, under the same
answer-key mappings. The eight models answer \MatchPaperTPR{} of the present
questions yes and raise \MatchPaperFPR{} false alarms on the absent ones. The
two arms share their acoustics but query different categories, so this rate is
not directly comparable to the accuracy in Table~\ref{tab:main}.
\subsection{Failures concentrate on specific source events}
\label{sec:events}

\begin{table}[tb]
\centering
\caption{Where the ranking failures live (\%). The \CatCount{} ESC-50
categories with $\ge$30 correct answers are split into quintiles by their
failure share (Q1 highest); its Q1/Q5 role sets are then re-scored on the
independently re-paired panel and re-rendered in the target control, which
covers only those two quintiles. Shares are over each panel's own correct
answers; the frozen row is that control's own baseline, not the main panel
restated. Gap is Q1$-$Q5, and its frozen-to-replaced
difference-in-differences is \SubstDiD{} points (\SubstDiDCI{}). The re-paired
row pools \CatRepModels{} models against \SubstModels{} elsewhere.}
\label{tab:category}
\small
\setlength{\tabcolsep}{3pt}
\renewcommand{\arraystretch}{0.97}
\begin{tabular}{lrrrrrr}
\toprule
Panel & Q1 & Q2 & Q3 & Q4 & Q5 & Gap \\
\midrule
Main (frozen targets) & 22.3 & 10.4 & 4.4 & 2.0 & 0.8 & $+21.5$ \\
\quad re-paired backgrounds & 31.2 & 20.3 & 12.0 & 7.8 & 5.3 & $+25.9$ \\
\quad targets frozen & 21.4 & -- & -- & -- & 0.5 & $+20.9$ \\
\quad targets replaced & 17.3 & -- & -- & -- & 1.6 & $+15.7$ \\
\bottomrule
\end{tabular}

\end{table}

Table~\ref{tab:category} shows that the failures concentrate on particular
source events. Ranking the categories once over all models pooled, the
highest quintile fails \CatQWOne{} of its correct answers against
\CatQWFive{} for the lowest, a gap of \CatGap{} points (95\% CI \CatGapCI{}).
The gap persists on the
independently re-paired panel (\CatRepGap{} points, \CatRepGapCI{}; the same
direction for all \CatRepModels{} models), and the per-category ordering
transfers across the two panels (Spearman $\rho=\CatRho$ over \CatShared{}
categories). Nor is it a rescaling of answer
confidence: the gap stays between \CatMarginLo{} and \CatMarginHi{} points
inside bins of the mixture margin $m_f$.
Both panels hold the target recording fixed, leaving the category entangled
with the possibility that these particular clips are simply
hard. So we re-rendered every Q1 and Q5 role with a different audited
recording of the same category, holding the background, SNR, overlap,
distractor pairing, question, and answer mapping byte-identical and adjusting
only the target waveform and its RMS-matched gain. The \SubstClips{}
replacement recordings fall unevenly over the \SubstRoles{} roles, giving
\SubstUnits{} evaluation units across the \SubstModels{} models
(\SubstConditions{} with the targets frozen); each frozen answer is reused for
every replacement of its role, giving both arms identical design-cell
multiplicities. The target-removed view is the same
file in both arms, and the gap narrows from \SubstGapFrozen{} to
\SubstGapSubst{} points (\SubstGapSubstCI{}), a difference-in-differences of
\SubstDiD{} points
(\SubstDiDCI{}), staying positive for
\SubstModelsOrdered{}/\SubstModels{} models (Table~\ref{tab:category}); the
per-role ordering transfers at Pearson $r=\SubstRoleRho$ over \SubstRoleN{} roles. Restricting to the units both
target versions answer correctly leaves \SubstGapSubstBoth{} points
(\SubstGapSubstBothCI{}) against \SubstGapFrozenBoth{} for the same units
frozen, so the surviving gap is not an artifact of the changing denominator.
The Q1--Q5 gap persists for the tested replacement recordings, although its
magnitude depends on target choice.

\subsection{Failures rise with acoustic competition}
\label{sec:stress}

\begin{table}[t]
\centering
\caption{Paired contrasts of the ranking-failure share (\%): $\Delta=$B$-$A in
points, with 95\% source-tuple bootstrap intervals. Competition rows use units
correct at every level; the one-kept-rival row restricts $\max_j$ to one
randomly kept rival against that cell's two one-distractor conditions; the
training row compares the AHA adapter with its base model on conditions both
answer correctly.}
\label{tab:contrasts}
\small
\setlength{\tabcolsep}{3.5pt}
\renewcommand{\arraystretch}{0.97}
\begin{tabular}{lrrrr}
\toprule
Contrast & A & B & $\Delta$ & 95\% CI \\
\midrule
\multicolumn{5}{l}{\emph{Acoustic competition (common-correct units)}}\\
Target SNR, $+5\to-5$~dB (5,466) & 4.9 & 8.4 & +3.5 & [2.7, 4.3] \\
\quad same, RMS-matched views & 5.2 & 8.0 & +2.9 & [2.1, 3.8] \\
Distractors, $1\to2$ (5,120) & 4.0 & 8.3 & +4.3 & [2.8, 5.9] \\
\quad same, one randomly kept rival & 4.0 & 6.5 & +2.5 & [1.3, 3.7] \\
\multicolumn{5}{l}{\emph{Training (Qwen2.5-Omni-7B $\to$ +AHA, common-correct)}}\\
Ranking-failure share & 3.1 & 3.1 & +0.0 & [-0.5, 0.4] \\
\bottomrule
\end{tabular}

\end{table}

Table~\ref{tab:contrasts} reports the share on units correct at
every level of the varied factor, so only the acoustics change. Across the
three target SNRs the share
climbs \SnrRankFailHigh{} $\to$ \SnrRankFailMid{} $\to$ \SnrRankFailLow{},
raising it from $+5$ to $-5$~dB by \SnrRankFailDelta{} points
([\SnrRankFailDeltaCI{}]), positive in all \SnrRankFailIncreasingModels{}
models, the smallest by \SnrRankFailMinModelLift{} points, though every answer
entering the comparison stays correct. Over all units at each SNR accuracy
falls from \SnrAccHigh{} to \SnrAccLow{}, and the
gradient survives the RMS-matched control rescaling each removal view to the
mixture's own loudness.

Added competition, not just added candidates, drives the rise. At a fixed
target-to-distractor SNR the distractors are scaled jointly, so two distractors
carry the same summed distractor power as one: doubling them at constant energy
still raises failures substantially. Scoring the same mixtures with one randomly
kept rival also rises clearly over the one-distractor baseline.

\subsection{The phenomenon survives re-paired backgrounds}
\label{sec:repair}

The failures are not an artifact of the distractor pairing. The main panel ties
each target role to one fixed pair, so we redrew both distractors for each of
the \ReplRoleCount{} frozen target roles, five times, on the full-overlap
two-distractor cells, holding the target clip, gain, and SNR fixed and drawing
new distractors blind to model outcomes. The share is \ReplRankFail{}, and on the
\ReplFrozenMatchedDenom{} answers correct under both pairings the frozen
pairing fails \ReplFrozenMatched{} against \ReplRepairMatched{} for the
re-paired backgrounds, a difference of \ReplMatchedDelta{} points
(95\% CI \ReplMatchedDeltaCI{}).

The split carries information beyond the binary flag. Of the \UtilTotalN{}
correct frozen-mixture answers the full check flags, \UtilConflictN{} carry an
ordering conflict and \UtilRepairableN{} a threshold-repairable error.
Conditional on a correct redraw, conflict answers fail \UtilConflictFail{} of
\UtilConflictRedrawN{} eligible redraws against \UtilRepairableFail{} of
\UtilRepairableRedrawN{}---\UtilRawDelta{} points, positive in all
\UtilModelsPositive{} models. Standardising those conflict answers within cells of
model, mixture margin ($0.5$-wide bins), and SNR, dropping cells under three repairable
units, each cell's repairable units setting the reference rate and conflict units
weighted by eligible redraws,
leaves \UtilMatchedDelta{} points (95\% CI \UtilMatchedDeltaCI{}) over
\UtilMatchedUsed{} redraws in \UtilCellCount{} strata, with a baseline unit's
five redraws counting as one unit in the source-tuple bootstrap. Restricting to
the \UtilSameMapUsed{} redraws with the original answer mapping gives
\UtilSameMapDelta{} points (95\% CI \UtilSameMapDeltaCI{}). This supports using
\method{} to prioritize correct answers for robustness testing.

Two controls support the reading. Detectors fixed by construction recover the
expected verdict, with $L(z)=\log(1+z/f)$ the panel's scoring map, $f$ a floor
at $10^{-3}$ of the loudest stem, and $E_i$ the RMS of stem $i$ rather than of
the mixture: \emph{oracle} $L(E_t)$ reads only the queried stem and
never fails, \emph{shortcut} $L(\max_j E_j)$ never reads the target and always
fails, and the \emph{RMS-sum} proxy $L(E_t+{\textstyle\sum_j} E_j)$, carrying no
source identity, lands at \ProbeLoudnessRankFail{}. On the
\AHACommonCorrectCount{} conditions where it and its base model both answer
correctly, the AHA adapter shows no detectable reduction in ordering conflicts
(Table~\ref{tab:contrasts}).
\section{Conclusion}
\label{sec:conclusion}
A correct presence answer can hide an inconsistent counterfactual judgment.
\method{} reports \RankFail{} ranking failures among \AttrCorrectCount{} correct
answers, clustering on particular source events and rising with acoustic
competition. Decomposing the flag separates ordering conflicts from
threshold-repairable errors and identifies correct answers at higher
ranking-failure risk under
background re-pairing, so accuracy alone cannot certify grounding in the sound a
model names.\newpage
\makeatletter
\renewcommand\section{\@startsection{section}{1}{\z@}%
  {-1.1ex plus -.2ex minus -.1ex}{0.4ex plus .1ex}{\normalfont\bfseries}}
\makeatother
\section{Acknowledgments}
No funding was received for this study; the authors have no relevant
financial or nonfinancial interests to disclose.
\section{Compliance with Ethical Standards}
Only public ESC-50 audio was used. No human or animal subjects were
involved: the two listeners were author-team members checking audio
fidelity, not study participants, so no ethical approval was required.
\makeatletter\def\@listi{\leftmargin\leftmargini\parsep 0pt\itemsep 0pt\topsep 0pt}\makeatother
\bibliographystyle{IEEEbib}
\bibliography{references}

\end{document}